\documentclass[11pt,showpacs,preprintnumbers,amsmath,amssymb,prd,nofootinbib,superscriptaddress]{revtex4-2}

\usepackage{dcolumn}
\usepackage{bm}
\usepackage{ifpdf}
\usepackage{hyperref}
\usepackage{bm}
\usepackage{xcolor,color,graphicx,graphics,physics}
\usepackage[spanish,english]{babel}
\usepackage[latin1]{inputenc}
\usepackage[OT1]{fontenc}
\usepackage{latexsym,amssymb,amsmath,amsfonts, slashed}
\usepackage{makeidx}
\usepackage{epsfig,subfigure}
\usepackage{natbib}
\usepackage{epstopdf}
\usepackage{mathrsfs}
\usepackage{hyperref}
\hypersetup{colorlinks=true, linkcolor=blue, citecolor=blue, urlcolor=blue}
\usepackage{enumerate}

\usepackage{braket}
\usepackage{fixmath}

\everymath{\displaystyle}
\usepackage{graphicx}

\usepackage[T1]{fontenc}
\usepackage{amsmath}
\usepackage{amssymb}
\usepackage{graphicx}
\usepackage{xcolor}

\newcommand{\bea}{\begin{eqnarray}}
\newcommand{\eea}{\end{eqnarray}}

\newcommand{\orcid}[1]{\href{https://orcid.org/#1}{\includegraphics[width=10pt]{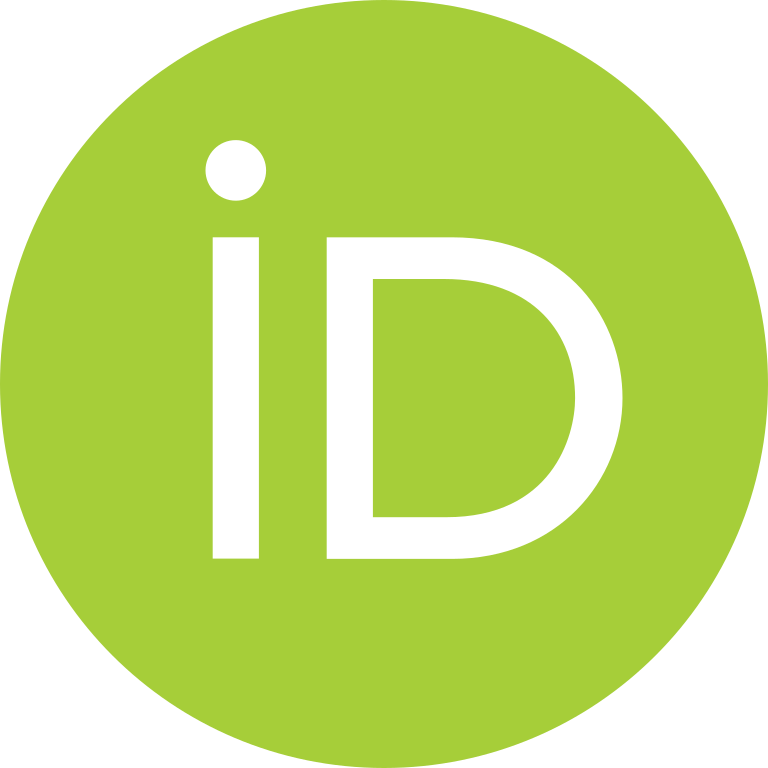}}}

\begin{document}

\title{Exploring regular black holes within the framework of the TFD formalism}

\author{Jhonny A. A. Ruiz \orcid{0000-0002-4384-2545}}
\email{jhonny.ruiz@ufjf.br}
\affiliation{Departamento de F\'{\i}sica, ICE,
Universidade Federal  de Juiz  de  Fora,\\
36036-100, Juiz de Fora, Minas Gerais, Brazil}

\author{A. F. Santos \orcid{0000-0002-2505-5273}}
\email{alesandroferreira@fisica.ufmt.br}
\affiliation{Instituto de F\'{\i}sica, Universidade Federal de Mato Grosso,\\
78060-900, Cuiab\'{a}, Mato Grosso, Brazil}

\begin{abstract}

In this paper, a massive scalar field non-minimally coupled with gravity is considered. Assuming a regular and spherically symmetric background, specifically a regular black hole solution, the energy-momentum tensor is calculated. Using the Thermo Field Dynamics (TFD) formalism and its topological structure, the corresponding Stefan-Boltzmann law and Casimir effect are investigated. The results for the regular black hole are compared with those obtained for the Schwarzschild black hole.

\end{abstract}

\maketitle

\section{Introduction}

Black hole (BH) physics is one of the most relevant research areas in Astrophysics and Cosmology today due to its fundamental relationships with Conformal Field Theory (CFT) and Quantum Field Theory (QFT) \cite{Hawking:1975vcx, Maldacena:1997re, Witten:1998qj}. It also has an experimental connection with the recent measurements of gravitational waves, which have profound theoretical and observational implications for the primordial power spectrum \cite{Planck2018VI, LIGOScientific:2016aoc, NANOGrav:2020bcs}. The presence of physical singularities in BH solutions for the vacuum Einstein field equations (EFEs) has been widely discussed, and many alternative models have been proposed to avoid central divergences, which are inherent, for instance, in the simple Schwarzschild BH \cite{Schwarzschild:1916uq, Hawking:1971vc, Bekenstein:1973ur, Berti:2009kk}. In this context, Regular Black Holes (RBHs) have appeared as a family of vacuum spherically symmetric solutions with no singularities, making them physically more interesting from both theoretical and experimental points of view \cite{bardeen1968non, Hayward:2005gi, Herdeiro:2015waa, Bambi:2013ufa}.

Besides vacuum BH and RBH solutions, it is possible to consider different matter components and even alternative gravitational theories to avoid divergences and provide a clearer interpretation of BH parameters \cite{Emparan:2008eg, Herdeiro:2014goa, Martinez:2005di, Bronnikov:2005gm}. In this context, the scalar field plays a remarkable role, as it provides an interesting approach to understanding some BH thermodynamical properties from a quantum perspective through the inclusion of temperature effects on its propagator using the Thermo Field Dynamics (TFD) formalism.

In TFD formalism, temperature and size effects are topologically introduced by doubling the Fock space. The doubled Fock space is composed of the standard Fock space $\mathcal{S}$ and  the dual (tilde) Fock space, defined as $\mathcal{\tilde{S}}$. Its main advantage lies in the geometrical interpretation of the Stefan-Boltzmann law and the Casimir effect as manifestations of the invariance of the Bogoliubov transformation in this doubled Fock space \cite{Umezawa1, Umezawa2, Book, Umezawa22, Khanna1, Khanna2, Santana1, Santana2}. Another key aspect of this formalism is that it allows the investigation of very different physical phenomena, such as thermal and size effects, on an equal footing.

In the present work, we evaluate the consequences of including temperature and size effects in a massive and non-minimal scalar field using the TFD formalism. For this development, a curved spacetime as given by an RBH metric will be considered. The physical consequences for the Stefan-Boltzmann law and the Casimir effect are calculated. The Bardeen and Hayward RBH solutions are compared with the Schwarzschild case.

The paper is organized as follows. In Sec.~\ref{Sec2}, the massive and non-minimal scalar field is presented. The energy-momentum tensor is calculated and the quantum propagator for this scalar field is given. The RBH is introduced as the gravitational background. In Sec.~\ref{Sec3}, TFD formalism is briefly introduced. A finite energy-momentum tensor is investigated. In Sec.~\ref{Sec4}, thermal effects for a regular black hole are discussed. In Sec.~\ref{Sec5}, size effects are introduced, and the Casimir effect is determined. Finally, in Sec.~\ref{Concl}, some concluding remarks are made.

\section{Massive scalar field}\label{Sec2}

In this section, the massive scalar field is introduced, and the energy-momentum tensor for this theory is calculated. Additionally, the RBH (Regular Black Hole) is presented as the gravitational background. Consider the Lagrangian density for a massive and non-minimal scalar field \cite{faraoni1998conformal, Copeland:2006wr, Capozziello:2011et, buchbinder2021introduction}
\begin{equation}
    \mathcal{L}=\sqrt{-g}\,\left[\frac{1}{2}\partial_{\mu}\phi\partial^{\mu}\phi-\frac{1}{2}m^2\phi^2+f(\phi)R\right],
\end{equation}
where $g$ is the metric determinant, $R$ the Ricci scalar and $f(\phi)$ is the non-minimal coupling function. The equation of motion associated with this Lagrangian density is
\begin{equation}
    \square\phi+m^2\phi-R\,f'(\phi)=0,
\end{equation}
with $ \square=g^{\mu\nu}\nabla_{\mu}\nabla_{\nu}$ and the prime denotes the derivative with respect to the field.

To investigate applications using the TFD formalism, the most important physical quantity to calculate is the energy-momentum tensor. The general expression for the energy-momentum tensor of this non-minimal scalar field theory is given by
\begin{equation}
    T_{\mu\nu}=\partial_{\mu}\phi\partial_{\nu}\phi-\frac{1}{2}g_{\mu\nu}\partial^{\kappa}\phi\partial_{\kappa}\phi+\frac{1}{2}g_{\mu\nu}m^2\phi^2+2\left(R_{\mu\nu}-\frac{1}{2}g_{\mu\nu}\,R+g_{\mu\nu}\square-\nabla_{\mu}\nabla_{\nu}\right)f(\phi).
\end{equation}
Assuming that the non-minimal coupling function $f(\phi)$ is given by $f(\phi)=\frac{1}{2}\xi\phi^2$, where $\xi$ is the conformal parameter, the energy-momentum tensor takes the form
\bea
    T_{\mu\nu}=\partial_{\mu}\phi\partial_{\nu}\phi-\frac{1}{2}g_{\mu\nu}g^{\kappa\lambda}\partial_{\kappa}\phi\partial_{\lambda}\phi\nonumber+\left[\xi\left(R_{\mu\nu}-\frac{1}{2}g_{\mu\nu}\,R+g_{\mu\nu}\square-\nabla_{\mu}\nabla_{\nu}\right)\textcolor{black}{+\frac{1}{2}g_{\mu\nu}m^2}\right]\phi^2.
\eea

To avoid divergences from the product of field operators at the same space-time point, this tensor is written at different space-time points as
\begin{align}
    T_{\mu\nu}(x)=&\lim_{x'\to x}\tau \Bigl\{\partial_{\mu}\phi(x)\partial'_{\nu}\phi(x')
    -\frac{1}{2}g_{\mu\nu}\partial^{\kappa}\phi(x)\partial'_{\kappa}\phi(x')\nonumber\\&+\left[\xi\left(R_{\mu\nu}-\frac{1}{2}g_{\mu\nu}\,R+g_{\mu\nu}\square-\nabla_{\mu}\nabla_{\nu}\right)\textcolor{black}{+\frac{1}{2}g_{\mu\nu}m^2}\right]\phi(x)\phi(x')\Bigr\},
\end{align}
where $\tau$ is the time ordering operator. Considering canonical quantization, the equal time commutation relation is
\begin{equation}
    [\phi(x), \partial'^{\mu}\phi(x')]=i\,n^{\mu}_{0}\,\delta(\vec{x}-\vec{x'}),
\end{equation}
and 
\begin{equation}
    \partial^{\rho}\,\theta(x^0-x'^0)=i\,n_0^\mu\,\delta(x^0-x'^0),
\end{equation}
with $n^\mu_0=(1, 0, 0, 0)$ as a unitary time-like vector, $\theta(x_0-x'_0)$ as the Heaviside step function and $\delta(\vec{x}-\vec{x'})$ as the usual $3-dimensional$ delta function \footnote{Here we are using the shorthand notation $x^{\mu}\equiv x=(x^0,\vec{x})$}. Thus, the energy momentum tensor becomes
\begin{equation}
    T_{\mu\nu}(x)=\lim_{x'\to x}\left\{\Gamma_{\mu\nu}\,\tau\left(\phi(x)\phi(x')\right)-I_{\mu\nu}\delta (x-x')\right\},
\end{equation}
where
\begin{equation}
 \Gamma_{\mu\nu}=\frac{1}{2}g_{\mu\nu}\partial^{\kappa}\partial'_{\kappa} -\partial_{\mu}\partial'_{\nu}+\xi\left(R_{\mu\nu}-\frac{1}{2}g_{\mu\nu}\,R+g_{\mu\nu}\square-\nabla_{\mu}\nabla_{\nu}\right)\textcolor{black}{+\frac{1}{2}g_{\mu\nu}m^2}  \label{8}
\end{equation}
and
\begin{equation}
    I_{\mu\nu}=-\frac{i}{2}\,g_{\mu\nu}n_{0}^{\kappa}n_{0\kappa}+i\,n_{0\mu}n_{0\nu}.
\end{equation}

In order to quantify the thermal and size effects, we need to calculate the vacuum expectation value of the energy-momentum tensor. This is given as
\begin{eqnarray}\label{Tavr}
   \langle T_{\mu\nu}(x)\rangle &=&\bra{0}T_{\mu\nu}(x)\ket{0}\nonumber\\
   &=&\lim_{x'\to x}\left[i\,\Gamma_{\mu\nu}\, G_0(x-x')-I_{\mu\nu}\,\delta(x-x')\right],
\end{eqnarray}
where
\begin{equation}
    i\,G_0(x-x')=\bra{0}\tau\left(\phi(x)\phi(x')\right)\ket{0},
\end{equation}
is the massive scalar field propagator, which can be written as
\begin{equation}\label{Gzerox}
    G_{0}(x-x')=-\frac{i\,m}{4\pi^2}\frac{K_{1}\left(m\sqrt{-(x-x')^2}\right)}{\sqrt{-(x-x')^2}}
\end{equation}
in coordinates space $x$, or 
\begin{equation}\label{kpropagator}
    G_0(k)=\frac{1}{k^2-m^2+i\epsilon},
\end{equation}
in momentum space $k$. Here $K_{\nu}$ is the modified Bessel function, which can be written as
\begin{equation}
    K_{\nu}(x)=i^{-\nu}\,J_{\nu}(ix)=e^{-\nu\pi\pi/2}J_{\nu}(xe^{i\pi/2})
\end{equation}
where $J_{\nu}$ are defined as the Bessel functions of the first kind \cite{gradshteyn2014table}.

Our main focus is to investigate regular black holes within this theory. A regular black hole (RBH) is a vacuum and spherically symmetric solution of Einstein field equations, whose line element can be expressed as follows
\begin{equation}\label{dsrbh}
    ds^2=-f(r)dt^2+\frac{dr^2}{f(r)}+r^2\left(d\theta^2+\sin^2\theta\,d\phi^2\right),
\end{equation}
where $f(r)$ is a general function of $r$ defined as
\begin{equation}\label{freg}
    f(r)=1-\frac{2M}{r\left[1+\left(r_0/r\right)^{q}\right]^{p/q}}.
\end{equation}
Here $M$ is the mass of the RBH, $q$ and $p$ are integers parameters and $r_0$ can be seen as the fundamental length of the RBH \cite{Bardeen:1973gs}. Such a line element reproduces known solutions for a suitable choice of $p$ and $q$, such as the Bardeen ($p=3$, $q=2$) and Hayward ($p=q=3$) solutions \cite{bardeen1968non, Hayward:2005gi}. Although this solution has no singularity at $r=0$, an event horizon does exist where ($f(r_h)=0$).

Considering this vacuum solution, our focus is to calculate the energy-momentum tensor. The main quantity we need to determine is $\Gamma_{\mu\nu}$, as defined in Eq. (\ref{8}). It takes the form
\begin{equation}
\Gamma_{\mu\nu}=\frac{1}{2}g_{\mu\nu}\partial^{\kappa}\partial'_{\kappa} -\partial_{\mu}\partial'_{\nu}+\xi\left(g_{\mu\nu}\square-\nabla_{\mu}\nabla_{\nu}\right)\textcolor{black}{+\frac{1}{2}g_{\mu\nu}m^2} 
\end{equation}
\noindent with its $00$ and $11$ components as
\begin{equation}\label{gamma00}
    \Gamma_{00}=-\frac{1}{2}\partial_0\partial_0-\left(\frac{1}{2}+\xi\right)f(r)\hat{D}\textcolor{black}{-\frac{1}{2}f(r)m^2}
\end{equation}
\begin{equation}\label{gamma11}
    \Gamma_{11}=-\frac{1}{2}\partial_1\partial_1-\left(\frac{1}{2}+\xi\right)f(r)^{-1}\hat{D}\textcolor{black}{+\frac{1}{2}f(r)^{-1}m^2}.
\end{equation}
Here, we have defined the differential operator
\begin{equation}
\hat{D}=f(r)\partial_1\partial_1+\frac{1}{r}\partial_2\partial_2+\frac{1}{r^2\sin{\theta}^2}\partial_3\partial_3.
\end{equation}
\noindent At this point, let us investigate some applications using the tools provided by the TFD formalism. We can calculate a Stefan-Boltzmann law associated with a scalar field having the RBH as the gravitational background. Similarly, the Casimir effect can be analyzed. In order to develop this study, the TFD formalism is introduced in the next section.

\section{Thermo Field Dynamics}\label{Sec3}

In this section, a thermal quantum field theory known as the TFD (Thermo Field Dynamics) formalism is briefly introduced. For a review of this formalism, see references \cite{Umezawa1, Umezawa2, Book, Umezawa22, Khanna1, Khanna2, Santana1, Santana2}. This formalism involves considering the vacuum expectation value of an arbitrary operator as equivalent to the statistical average. To achieve this, a new ground state, called the thermal vacuum, must be proposed. Thus, the average value of an arbitrary operator $\mathcal{O}$ in a thermal vacuum is defined as
\begin{equation}
    \braket{\mathcal{O}}=\braket{0(\beta)|\mathcal{O}|0(\beta)}
\end{equation}
where $\beta=1/(k_{B}T)$, with $k_{B}$ being the Boltzmann constant and $T$ the temperature. This construction requires two ingredients: (i) the doubled Fock space $\mathcal{S}_T$ given by the tensor product
\begin{equation}
    \mathcal{S}_T=\mathcal{S}\otimes\mathcal{\tilde{S}},
\end{equation} 
where $\mathcal{S}$ is the standard Fock space and $\mathcal{\tilde{S}}$ is the dual (tilde) space. (ii) The Bogoliubov transformation, which involves a rotation between $\mathcal{S}$ and $\mathcal{\tilde{S}}$, in the form
\begin{equation}
    \left(\begin{array}{c}
         \mathcal{O} (k,\alpha)  \\
         \eta \tilde{\mathcal{O}}^{\dag}(k,\alpha)
    \end{array}\right)=\mathcal{B}(\alpha) \left(\begin{array}{c}
         \mathcal{O} (k)  \\
         \eta \tilde{\mathcal{O}}^{\dag}(k)
    \end{array}\right),
\end{equation}
where $\eta=1,-1$, for fermions and bosons, respectively,  $\mathcal{B}(\alpha)$ is the Bogoliubov factor defined as 
\begin{equation}
   \mathcal{B}(\alpha)=\left(\begin{array}{cc}
         u(\alpha) & -v(\alpha)  \\
         \eta\,v(\alpha) & u(\alpha)
    \end{array}\right)
\end{equation}
and
\begin{equation}
    v^{}(\alpha)=\left(e^{\alpha\omega_k}-1\right)^{-1},\qquad\qquad u^{2}(\alpha)=1+v^{2}(\alpha)
\end{equation}
which satisfy the identity
\begin{equation}
    u^{2}(\alpha)+\eta\,v^{2}(\alpha)=1,
\end{equation}
with $\alpha=(\alpha_0,\alpha_1,...,\alpha_{D-1})$ as the compactification parameter over the $D$-dimensions and $\omega_k=k_0$ as the associated energy
 \cite{Book}. 

Another characteristic of the TFD formalism is the introduction of a compactification parameter in the propagator of the theory. Here, we introduce this parameter in the scalar field propagator \cite{Santana2}, starting with the expression
\begin{eqnarray}
    G_{0}^{AB}(x-x';\alpha)&=&i\braket{0(\alpha)|\tau\left[\phi^{A}(x)\phi^{B}(x')\right]|0(\alpha)}\nonumber\\
    &=&i\int{\frac{d^{4}x}{(2\pi)^{4}}e^{-ik(x-x')}G_{0}^{AB}(k;\alpha)}
\end{eqnarray}
where
\begin{equation}
    G_{0}^{AB}(k;\alpha)=\mathcal{B}^{-1}(\alpha)G_{0}^{AB}(k)\mathcal{B}(\alpha),
\end{equation}
with
\begin{equation}
    G_{0}^{AB}(k)=\left(\begin{array}{cc}
         G_{0}(k) & 0  \\
         0 & \eta G_{0}^{*}(k)
    \end{array}\right)
\end{equation}
and $G_{0}(k)$ is the scalar field propagator defined in \eqref{kpropagator}\footnote{Here A, B=1,2 represent the doubled space and $G_{0}^{*}(k)$ is the complex conjugate of $G_{0}(k)$.}. Its physical component is given by the component $A=B=1$ \cite{Khanna1}, such that
\begin{equation}
    G_{0}^{AB}(k;\alpha)=G_{0}(k)+\eta\,v^{2}(k;\alpha)\left[G_{0}^{*}(k)-G_{0}(k)\right],
\end{equation}
where $v^{2}(k;\alpha)$ is the generalized Bogoliubov transformation \cite{Khanna2}.

In this context, let's rewrite the vacuum expectation value of the energy-momentum tensor, given in Eq. (\ref{Tavr}), using the doubled notation. Thus,
\begin{eqnarray}
   \langle T_{\mu\nu}^{(AB)}(x;\alpha)\rangle =\lim_{x'\to x}\left[i\,\Gamma_{\mu\nu}\, G_0^{(ab)}(x-x';\alpha)-I_{\mu\nu}\,\delta(x-x')\delta^{AB}\right].
\end{eqnarray}
To obtain a physical energy-momentum tensor that leads to finite quantities, a renormalization procedure is necessary. Here, the Casimir prescription is applied. The resulting finite energy-momentum tensor is given by
\bea
{\cal T}_{\mu\nu (ab)}(x;\alpha)=\langle T_{\mu\nu(ab)}(x;\alpha)\rangle-\langle T_{\mu\nu(ab)}(x)\rangle.
\eea
With this procedure a measurable physical quantity is obtained. Explicitly
\bea
{\cal T}_{\mu\nu (ab)}(x;\alpha)=i\lim_{x'\rightarrow x}\left\{\Gamma_{\mu\nu}(x,x')\overline{G}_0^{(ab)}(x-x';\alpha)\right\},\label{VEV}
\eea
where 
\bea
\overline{G}_0^{(ab)}(x-x';\alpha)=G_0^{(ab)}(x-x';\alpha)-G_0^{(ab)}(x-x').
\eea

In the next sections, Eq. (\ref{VEV}) is used to explore the topological structure of the TFD formalism, and then some applications are investigated.


\section{Thermal effects for a regular black hole}\label{Sec4}

The TFD formalism represents a field theory defined on the topology $\Gamma_D^d=(\mathbb{S}^1)^d\times \mathbb{R}^{D-d}$ where $1\leq d \leq D$. In this context, 
$D$ denotes the total number of space-time dimensions, while $d$ signifies the number of dimensions that have been compactified. This formalism allows for the compactification of any subset of dimensions within the manifold $\mathbb{R}^{D}$. In this construction, the circumference of each $nth$ $\mathbb{S}^1$ in the topology $ \Gamma_D^d$ is specified by $\alpha_n$. First, let's consider the specific case  $\Gamma_4^1=\mathbb{S}^1\times\mathbb{R}^{3}$, where $\alpha=(\beta,0,0,0)$. In this case the time-axis is compactified in $\mathbb{S}^1$, with circumference $\beta$. For this case, thermal effects are introduced, and then the Stefan-Boltzmann law can be calculated.

In order to utilize the energy-momentum tensor expression, let's outline the generalized Bogoliubov transformation for this scenario
\bea
v^2(\beta)=\sum_{l_0=1}^{\infty}e^{-\beta k^0l_0}.\label{BT1}
\eea
Then the Green function is given by
\bea
\overline{G}_0(x-x';\beta)=2\sum_{l_0=1}^{\infty}G_0(x-x'-i\beta l_0n_0),\label{GF1}
\eea
where $n_0=(1,0,0,0)$ and $\overline{G}_0(x-x';\beta)\equiv \overline{G}_0^{(11)}(x-x';\beta)$, that is the physical component. Thus Eq. (\ref{VEV}) becomes
\begin{equation}\label{tmunu}
    \mathcal{T}_{\mu\nu(11)}(x;\beta)=2i\lim_{x'\to\,x}\left\{\Gamma_{\mu\nu}\sum_{l_0=1}^{\infty}G_0(x-x'-i\beta\,l_0\,n_0)\right\}.
\end{equation}

The propagator for the massive scalar field takes the form
\begin{equation}
     G_{0}(x-x'-i\beta\,l_0\,n_0)=-\frac{i\,m}{4\pi^2}\frac{K_{1}
     \left(m\sqrt{-(x-x'-i\beta\,l_0\,n_0)^2}\right)}{\sqrt{-(x-x'-i\beta\,l_0\,n_0)^2}}
\end{equation}
with
\begin{equation}
    (x-x'-i\beta\,l_0\,n_0)^2=-f(r)(t-t'-i\beta\,l_0\,n_0)^2+\frac{1}{f(r)}(r-r')^{2}+r^2(\theta-\theta')^2+r^2\sin{\theta}^2(\phi-\phi')^2,
\end{equation}
in correspondence to the RBH line element defined in Eq. \eqref{dsrbh}.

Using these ingredients and Eq. \eqref{gamma00}, the energy for the massive scalar field in the regular black hole geometry is given by the $00$ component of Eq. \eqref{tmunu}. After some calculations, the energy can be written as
\begin{eqnarray}\label{40E}
    E=-\frac{1}{4 \pi ^2 }\sum_{l_0=1}^{\infty}\left\{\frac{6m^2(1+\xi)}{l_0^2\beta^2}K_0[f(r)]+\frac{12m(1+\xi)}{l_0^3\beta^3\sqrt{-f(r)}}K_1[f(r)]\right\},
\end{eqnarray}
where we have defined
\begin{equation}
    K_{\nu}[f(r)]= K_{\nu}\left[m l_0 \beta \sqrt{- f(r)}\right],\qquad\nu=0,1,
\end{equation}
with $K_{\nu}[f(r)]$ being the Bessel functions and $\mathcal{T}^{00(11)}(x;\beta)\equiv E$. This represents the Stefan-Boltzmann law for the massive scalar field in a gravitational background described by a regular black hole. It is not possible to perform the sum in Eq. (\ref{40E}) and obtain a complete expression for the energy. To derive a closed-form expression for the energy in this context, let us consider the massless case. In this limit, the energy takes the simple form
\begin{equation}
    E=\frac{\pi^2(1+\xi)}{30\beta^4\,f(r)}.
\end{equation}
Thus, when $f(r)$ takes the usual Schwarzschild form $f(r)=1- 2M/r$, we recover the energy for a massless scalar field at finite temperature, as reported in previous work \cite{Ulhoa:2020vmn}, i.e.,
\bea
 E_{Sch}=\frac{\pi^2(1+\xi)}{30\beta^4}\left(\frac{r}{r-2M}\right).
\eea

On the other hand, the corresponding Bardeen and Hayward energies for the massless case are given by
\begin{equation}
    E_B=\frac{\pi^2(1+\xi)}{30\beta^4}\left\{1-\frac{2M}{r\left[1+\left(r_0/r\right)^2\right]^{3/2}}\right\}^{-1}
\end{equation}
\begin{equation}
    E_H=\frac{\pi^2(1+\xi)}{30\beta^4}\left\{1-\frac{2M}{r\left[1+\left(r_0/r\right)^3\right]}\right\}^{-1}.
\end{equation}

\begin{figure}[t]
    \centering
    \includegraphics[scale=1]{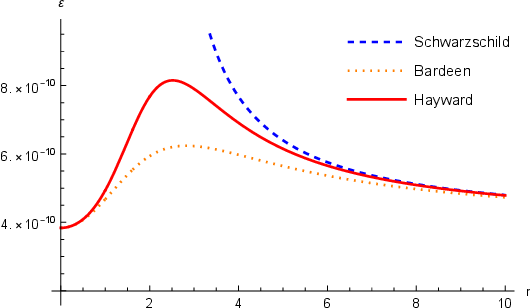}
    \caption{Comparing the energy (${\cal E}$) versus radius ($r$) for different gravitational background: Schwarszchild black hole (dashed), Bardeen regular black hole (dotted) and Hayward regular black hole (solid). Here, for simplicity, we have used ${\cal E}=E\times 10^{-9}$, $M=T=1$, $r_0=2$ and $\xi=1/6$. }
    \label{fig1}
\end{figure}

It is important to emphasize that these results show that the Stefan-Boltzmann law associated with the Schwarzschild black hole behaves differently when compared to the Bardeen and Hayward regular black holes. In Figure \ref{fig1}, the energy is shown for these three cases. The results demonstrate that the singularity due to the Schwarzschild solution persists in the energy, while for the Bardeen and Hayward solutions, the energy is always finite. In addition, it is interesting to note that, for large radii, all three cases take the same shape.

\section{Casimir effect at zero temperature}\label{Sec5}

To investigate size effects in TFD formalism, let's consider the topology $\Gamma^{1}_{4}=\mathbf{S^1}\times\mathbf{R^3}$ and the compactification parameter $\alpha$ defined as $\alpha=(0, i2b, 0, 0)$, where $2b$ is the circumference of $\mathbf{S^1}$. For this case,  the Bogoliubov transformation takes the form
\begin{equation}
v^2(b)=\sum_{l_1=1}^{\infty}e^{-i2bk^{1}l_1}
\end{equation}
and the Green function is
\begin{equation}
    \bar{G}_0(x-x';b)=2\sum_{l_1=1}^{\infty}G_0(x-x'-2bl_{1}n_1)
\end{equation}
with $n_1=(0, 1, 0, 0)$. Then the Casimir energy is calculated from the $00$ component of the energy-momentum tensor, which is given as
\begin{equation}
    \mathcal{T}_{00}^{11}(x;b)=2i\lim_{x'\to\,x}\left\{\Gamma_{00}\sum_{l_1=1}^{\infty}G_0(x-x'-i2b\,l_1\,n_1)\right\}.
\end{equation}
Using Eq. \eqref{gamma00} and defining  $ \mathcal{T}_{00}^{11}(x;b)=E_c$, the Casimir energy is obtained as
\begin{eqnarray}\label{casenzerot}
E_c&=&\sum_{l_1=1}^{\infty}\frac{m}{8\pi^2b^2 l_1^2}\Bigl\{\left[(1 + \xi){f(r)}^2 + b l_1 (1 + 2 \xi) f'(r)\right]mK_0(r)\nonumber\\
&+&\left[-2b^2l_1^2 f(r)m^2\xi+(1+\xi){f(r)}^2-bl_1 f'(r)(1+2\xi)(b^2l_1^2m^2-f(r))\right]\frac{\sqrt{-f(r)}}{bl_1}K_1(r)\Bigr\}.
\end{eqnarray}

To find the Casimir pressure, we use now
\begin{equation}
    \mathcal{T}_{11}^{11}(x;b)=2i\lim_{x'\to\,x}\left\{\Gamma_{11}\sum_{l_1=1}^{\infty}G_0(x-x'-i2b\,l_1\,n_1)\right\},
\end{equation}
which yields the expression at zero temperature
\begin{eqnarray}\label{casprezerot}
    P_c&=&\sum_{l_1=1}^{\infty} \frac{m}{8\pi^2b^3l_1^3}\left(\frac{\sqrt{-f(r)}}{f(r)}\right)^3\Bigl\{bl_1\frac{f(r)}{\sqrt{-f(r)}}\left[3(1+\xi)f(r)+bl_1f'(r)\right]m K_0(r)\nonumber\\
    &+&\left[3(1+\xi){f(r)}^2-b^3l_1^3m^2f'(r)+bl_1f(r)\left(f'(r)-bl_1m^2\right)\right]K_1(r)\Bigr\},
\end{eqnarray}
where Eq.  \eqref{gamma11} has been used. Here, we have defined $\mathcal{T}_{11}^{11}(x;b)=P_c$ and
\begin{equation}
        K_\nu(r)=K_\nu\left(\frac{2mbl_1}{\sqrt{- f(r)}} \right),\qquad\qquad \nu=0,1.
\end{equation}

In the case of zero mass and assuming that the gravitational background is described by a Schwarzschild black hole,  these expressions for the Casimir-Schwarzschild energy and pressure, reduce to
\begin{equation}
    E_{CS}=-\frac{(r-2M)^2}{1440\pi^2b^4r^4}\left[\pi^4r(r-2M)(1+\xi)-180bM(1+2\xi)\zeta(3)\right]
\end{equation}
and
\begin{equation}
    P_{CS}=-\frac{\pi^4(1+\xi)r(r-2M)+60bM\zeta(3)}{480b^4\pi^2r^2}.
\end{equation}
These results are in agreement with the previous paper \cite{Ulhoa:2020vmn}.

From the general expressions \eqref{casenzerot} and \eqref{casprezerot}, and in the massless limit for the scalar field, it is also possible to write down the Casimir-Bardeen and Casimir-Hayward energies and pressures, which are given by
\begin{equation}
    E_{CB}=\frac{1}{1440b^4\pi^2}f_B^2(r)\left[-\pi^4f_B(r)(1+\xi)-180bM(1+2\xi)\zeta(3)\frac{r^3-2rr_0^2}{\left(r^2+r_0^2\right)^{3/2}}\right]
\end{equation}
\begin{equation}
    E_{CH}=\frac{1}{1440b^4\pi^2}f_H^2(r)\left[-\pi^4f_H(r)(1+\xi)-180bM(1+2\xi)\zeta(3)\frac{Mr^4-2Mr_0^3}{\left(r^3+r_0^3\right)^{2}}\right]
\end{equation}
for the energies, and
\begin{equation}
    P_{CB}=-\frac{\pi^2(1+\xi)}{480b^4}+\frac{\pi^2M(1+\xi)r^2}{240b^4(r^2+r_0^2)^{3/2}}-\frac{Mr(r^2-2r_0^2)\zeta(3)}{8b^3\pi^2(r^2+r_0^2)^{5/2}}
\end{equation}
\begin{equation}
P_{CH}=-\frac{\pi^2(1+\xi)}{480b^4}+\frac{2Mr\left[\pi^4r(r^3+r_0^3)(1+\xi)-30b(r^3-2r_0^3)\zeta(3)\right]}{480b^4\pi^2(r^3+r_0^3)^2}
\end{equation}
for the pressures. Here we have used the notation 
\begin{equation}
    f_B(r)=1-\frac{2M}{r\left[1+(r_0/r)^2\right]^{3/2}},\qquad\qquad f_H(r)=1-\frac{2M}{r\left[1+(r_0/r)^3\right]}.
\end{equation}

\begin{figure}[t]
    \centering
    \includegraphics[scale=1]{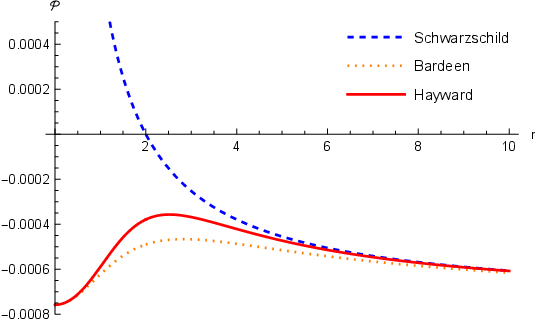}
    \caption{Casimir pressure (${\cal P}$) versus radius ($r$) for different cosmological backgrounds: Schwarzschild black hole (dashed), Bardeen regular black hole (dotted), and Hayward regular black hole (red). For simplicity, ${\cal P} = P_{BH} \hbar c$, where $P_{BH} = P_{CS}, P_{CB}, P_{CH}$. Additionally, it is assumed that $M = 1$, $b = 10^{-6}$, $r_0 = 1$, and $\xi = 1/6$. }
    \label{fig2}
\end{figure}

In order to understand the differences between these three solutions, the Casimir pressure has been plotted in Figure \ref{fig2}. The pressure associated with a Schwarzschild black hole exhibits an infinite value at the Schwarzschild radius. This radius marks the transition between positive and negative pressure, or repulsive and attractive force. On the other hand, the Bardeen and Hayward solutions show finite pressure for all values of $r$. The pressure in both cases is always negative, meaning the force is only attractive for all values of $r$.  For large radii, the pressures in the Schwarzschild, Bardeen, and Hayward solutions exhibit the same behavior, tending towards the same negative value.

\section{Conclusions}
\label{Concl}

Black hole physics have been fascinating since the beginning of general relativity. The singularities inherent in their physics present a challenge that has drawn significant attention. To study the physics of black holes without singularities, regular black hole solutions such as the Bardeen and Hayward solutions have been proposed. In this work, a massive scalar field non-minimally coupled with gravity is considered. Using the topological structure of the TFD formalism, thermal and size effects are investigated, considering a regular black hole as the gravitational background. The Stefan-Boltzmann law and the Casimir effect for the massive scalar field in curved spacetime are calculated. Our results show that these quantities exhibit different behavior when different regular black holes are considered. We compare the results for regular black holes with those obtained for the Schwarzschild black hole. While the results for the Schwarzschild black hole exhibit singularities, the results for the Bardeen and Hayward black holes are always finite. One of the most important results of this paper is related to the Casimir pressure. Here, three different solutions are analyzed. When the Schwarzschild spacetime is considered, there is a transition from attractive to repulsive force. Additionally, for $r$ smaller than the event horizon, the force is repulsive and goes to an infinite value. For the Bardeen and Hayward black holes, the force is always attractive and finite.

\section*{Acknowledgments}

This work by A. F. S. is partially supported by National Council for Scientific and Technological Development - CNPq project No. 312406/2023-1. J. A. A. Ruiz would like to thanks to Universidad Tecnol\'ogical del Uruguay and Universidade Federal de Juiz de Fora for the support during the development of this research.

\global\long\def\link#1#2{\href{http://eudml.org/#1}{#2}}
 \global\long\def\doi#1#2{\href{http://dx.doi.org/#1}{#2}}
 \global\long\def\arXiv#1#2{\href{http://arxiv.org/abs/#1}{arXiv:#1 [#2]}}
 \global\long\def\arXivOld#1{\href{http://arxiv.org/abs/#1}{arXiv:#1}}

\end{document}